\documentclass[%
 reprint,
superscriptaddress,
 amsmath,amssymb,
 prapplied,
noeprint
]{revtex4-2}

\usepackage{xcolor}
\usepackage{graphicx}% Include figure files
\usepackage{dcolumn}% Align table columns on decimal point
\usepackage{bm}% bold math
\usepackage[format=plain,justification=Justified]{caption}
\usepackage{subcaption}
\usepackage{gensymb}
\usepackage{float}
\usepackage{csquotes}

\usepackage[T1]{fontenc}
\usepackage{dirtytalk}

\usepackage{amsmath}
\DeclareMathOperator*{\argmax}{arg\,max}

\begin{document}

\preprint{APS/123-QED}

\title{Machine learning optimal control pulses in an optical quantum memory experiment}% Force line breaks with \\

\author{Elizabeth Robertson} 
 \email{elizabeth.robertson@dlr.de}
\author{Luisa Esguerra}
\affiliation{%
  Deutsches Zentrum für Luft- und Raumfahrt e.V. (DLR), Rutherfordstr. 2, 12489 Berlin, Germany
}%

\affiliation{%
  Technische Universität Berlin, Institute for Optics and Atomic Physics, Hardenbergstr. 36, 10623 Berlin, Germany
}%
\author{Leon Messner}
\affiliation{%
  Deutsches Zentrum für Luft- und Raumfahrt e.V. (DLR), Rutherfordstr. 2, 12489 Berlin, Germany
}%
\author{Guillermo Gallego}
\affiliation{%
   Einstein Center Digital Future and Science of Intelligence Excellence Cluster 10117 Berlin, Germany
}%
\author{Janik Wolters}
\affiliation{%
  Deutsches Zentrum für Luft- und Raumfahrt e.V. (DLR), Rutherfordstr. 2, 12489 Berlin, Germany
}%
\affiliation{%
  Technische Universität Berlin, Institute for Optics and Atomic Physics, Hardenbergstr. 36, 10623 Berlin, Germany
}%

\date{\today}% It is always \today, today,

\begin{abstract}
Efficient optical quantum memories are a milestone required for several quantum technologies including repeater-based quantum key distribution and on-demand multi-photon generation. We present an efficiency optimization of an optical electromagnetically induced transparency (EIT) memory experiment in a warm cesium vapor using a genetic algorithm and analyze the resulting waveforms. The control pulse is represented either as a Gaussian or free-form pulse, and the results from the optimization are compared. We see an improvement factor of 3(7)\% when using optimized free-form pulses. By limiting the allowed pulse energy in a solution, we show an energy-based optimization giving a 30\% reduction in energy, with minimal efficiency loss.
%\begin{description}
%\end{description}
\end{abstract}

\maketitle

%\tableofcontents

\section{\label{sec:level1} Introduction}
Optical memories have been long recognized as a required technology in the implementation of different quantum protocols, notably on-demand multi-photon generation \cite{Nunn.2008}, repeater-based quantum key distribution (QKD) \cite{Sangouard.2011, Gundogan.2021, Wallnofer.2022, Mol.2023}, and the translation of flying to stationary qubits, among others \cite{Manukhova.2017, Gundogan.2021b}. Indeed, the aforementioned protocols have highly efficient operation as a requirement, and often the memory efficiency bounds the attainable end-to-end efficiency of the system. In QKD implementations, for example, this leads to a logarithmic scaling in the time required for entanglement distribution \cite{Gundogan.2021}. Moreover, optical memories find applications in classical analog computing, in systems such as reservoir computing, where the efficiency of the optical memory provides a bound on the memory capacity of the reservoir \cite{Jaurigue.2021}. 

Similarly to the variety one sees in the \enquote{qubits zoo}, quantum memories are equally diverse in their form \cite{Lei.2023}. Spanning from single atoms \cite{Specht.2011}, to solid-state devices \cite{Gundogan.2015, Yang.2018}, a range of optical quantum memories \cite{Pu.2017} have been investigated providing a wide range of efficiencies. The most efficient demonstrated memory systems are those operating at the ultra-cold regimes. By removing several sources of decoherence, efficiencies of 75\% - 85\% \cite{Yang.2016, Bao.2012} have been achieved. However, solid state devices also boast high-efficiency operation, shown between 56\%  and 69\% \cite{Sabooni.2013, Hedges.2010}. 
\begin{figure*}
    \centering
    \includegraphics[width = \linewidth]{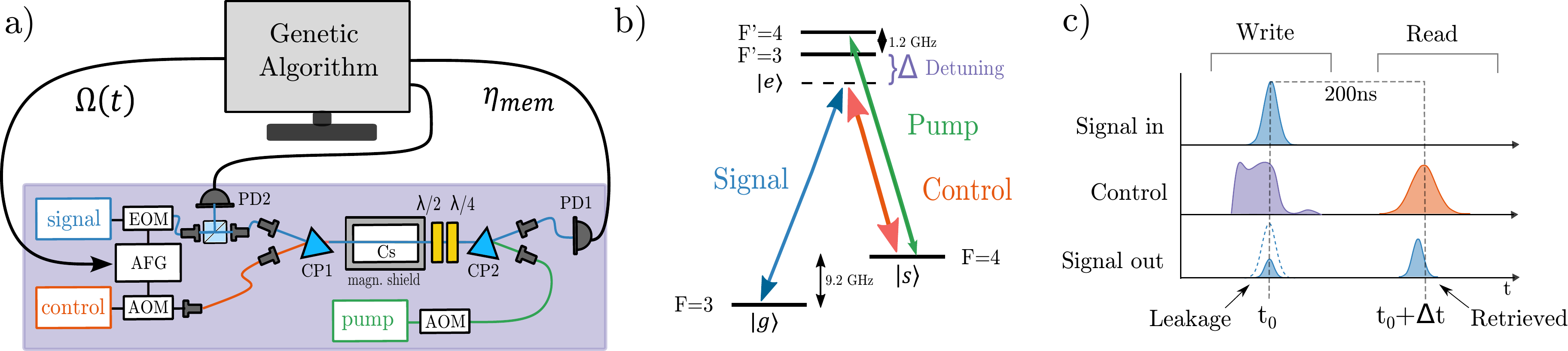}
    \captionof{figure}{ Experimental schematics. a) An overview of online genetic algorithm optimization setup. Possible solutions (control pulses) are generated by the genetic algorithm and are evaluated by carrying out a memory experiment. EOM: electro-optic modulator; AOM: accousto-optic Modulator; AFG: arbitrary function generator; CP1, CP2: calcite prisms; PD: photodiode; Cs: cesium vapor cell. b) The three-level $\Lambda$ system used in the memory protocol. The signal (control) lasers are red-detuned by 1GHz from their resonant hyper-fine transitions. A resonant pump laser populates the ground state $|g\rangle$. c) A typical memory experiment sequence. We denote $t_{\rm{0}}$ = 0 as the center of the signal pulse to be stored, and we read out such that the time between the signal in and the control read pulse is $\Delta t = 200\mathrm{ns}$. }
    \label{fig:1}
\end{figure*}
Warm vapor atomic memories based on electromagnetically induced transparency (EIT) have been highlighted as one of the most promising of these systems, as they are technologically simple, can be multiplexed \cite{Mener.2023}, and have been modeled to have high operating efficiency \cite{Gorshkov.2008}. Using strong magnetic control of atomic ensembles an 82\% internal efficiency at room temperature was demonstrated \cite{Guo.2019}.
Moreover, warm vapor memories have been shown to have an acceptance bandwidth of 0.66 GHz, making them suitable for interfacing with semiconductor quantum dots and other single photon sources \cite{Wolters.2017}.

%As strong magnetic control is impractical in remote deployment i.e. on a satellite, we further explore the optimization of warm atomic memories suited for space . 
%The optimization of warm atomic memories has been explored in both theory and experiment.
%In this report we present the optimization of warm atomic vapor memory, a scheme which has been highlighted for its possible suitability as a satallite based quantum repeater.
%and obtain the optimal spin wave within the optical medium. Here the
A  theoretical formulation and optimization of the three-level $\Lambda$ EIT scheme in free space was presented in Refs. \cite{Gorshkov.2007b, Gorshkov.2008}. There, they use a gradient ascent approach to optimize the control field,  using an analytically calculable gradient derived from the atomic spin wave, to learn the temporal shape of the optimal optical pulse. These simulation-optimized pulses were then transferred to the experiment, where they were shown to perform well \cite{Novikova.2008, Phillips.2008}. However, the model presented in these works is limited in the physics it accounts for; indeed the presence of a fourth level changes the shape of the optimal control pulses, as shown in Ref. 
\cite{Rakher.2013}. Moreover, there are further experimental effects such as four-wave mixing (FWM) and the inhomogeneous broadening of the excited state that are not accounted for in the four-level system models, and the optimal operating conditions of a memory experiment.
\\
Using the genetic algorithm, we learn optimal control pulses of a warm cesium vapor quantum memory, where the temporal shape is encoded either as a Gaussian with the amplitude, pulse width, and delay as free parameters, or a 16-parameter free-form pulse. 
In this contribution, we apply an optimization process to the experiment as a whole. In-experiment optimizations boast the benefit of accounting for a variety of experimental effects; device-specific transfer functions of optical modulators, inter-system delays between signal and control, and physical effects not captured in model systems.
However, as the spin-wave gradients are not accessible in an experimental setting, we must consider an alternative, gradient-free learning approach. Genetic algorithms are widely acknowledged as noise-robust gradient-free optimization algorithms, making them attractive for use in a wide range of atomic (optics)  experiments \cite{Hornung.2000, Zeidler.2001, Judson.1992, Gregoric.2017}.

%We present results supporting the conclusions presented in the theoretical works of Shinbrough and colleagues \cite{Shinbrough.2021};
 We find that the efficiency of memory experiments carried out with Gaussian pulses is similar to those of free-form pules, in the device-resolvable range of the experiment. This is confirmed in theory, albeit for the storage of shorter signals \cite{ Shinbrough.2021}.  We show experimentally that this trend holds for larger pulse widths; that is, we find that the use of free-from pulses provides an average improvement  of 3(7)\% for signals ranging between $\tau_{\rm{FWHM}} \gamma  = \{1.9,21.5\}$. Here, $\tau_{\rm{FWHM}}$ is the full-width-half-maximum (FWHM) of the signal pulse, given in units of $\tau$, where $\tau$ is $1/\gamma$ and $\gamma$ is the excited state decay rate.
Moreover, we demonstrate that temporal regions of particular importance to the efficiency when using free-form pulses are similar to those temporally overlapping with the signal pulse \cite{Shinbrough.2023}. 
We also illustrate the possibility of learning optimal pulses under further objective constraints, such as total pulse energy minimization. Here we show that we can reduce the energy of the learned signal and control pulses by 30\% with a minimal trade-off of 4(6)\% in efficiency. This finding might have important implications for reducing the readout noise - which is a well-known issue in warm vapor memories.

This work is structured as follows: In section \ref{sec: mem_optimize}, we first present the optical memory (Section \ref{sec:optMem}) and then the genetic algorithm (Section \ref{sec:Genetic_alg}). The analysis of the results is divided into two sections, first we consider the optimization of different-width signals stored (Section \ref{sec: results}). In Section \ref{sec:power_opt}, we discuss the results of the energy optimization. A brief discussion of improvements to the method and an outlook for further research are given in the final section.

%However, to optimize on and experiment requires a different learning approach as we cannot obtain gradients of the spin wave, that is a gradient descent appproach is not possible. 
%from molecular state excitation  to femto-second pulse generation controlling field ionization pathway of a Rydberg electron with a GA, to , naming but a few.
 %.
%\cite{Weinacht.1999} Pulse control with AOM to excite Cs, manipulate the shape of an atom's wavefunction - Not with GA, just with some feedback algorithm
%Ref. details a different approach to iterative in-experiment optimization; learning the shape of the incoming signal pulse, converging in as few as 3 iterations. This approach is not applicable for the control pulse as.... 

\section{\textbf{\label{sec: mem_optimize} Optical quantum memory optimization}}

\subsection{ \label{sec:optMem} Warm vapor memory}
%We wish to optimize the internal memory efficiency of an optical quantum memory, which utilizes EIT to mediate the storage of photons in collective spin states of cesium atoms.  B
The experimental setup we optimize has been described extensively in Ref. \cite{Esguerra.2023}; Figure \ref{fig:1}a) depicts a compressed setup and below we emphasize the differences in the setup used in this work. 

We use two lasers with linear, orthogonal polarization as the signal (S) and control (C) lasers, which are offset-locked to a frequency difference of 9.2 GHz. The signal and control pulses are generated with an electro-optic modulator and an accousto-optic modulator, respectively. This enables us to reach peak pulse powers of $P_{\mathrm{C, cw}}= 12.9\ \mathrm{mW}$, without the semiconductor optical amplifier and spontaneous emission filtering system used in Ref. \cite{Esguerra.2023}. A pump laser with power $ P _{\mathrm{P,cw}}= 15.1\ \mathrm{mW}$ is locked on the $F = 4 \rightarrow F' = 4$ transition, and counter propagates to the signal and control lasers. We red-detune both lasers to be $1\ \mathrm{GHz}$ off the atomic transition; a level scheme indicating the laser frequencies is shown in Fig.\ref{fig:1} b). The wavelength of the control laser is monitored using a wavemeter and is locked to it using a simple feedback loop, to prevent frequency drifts during the algorithm evolution. For each memory experiment carried out during the genetic algorithm evolution, we pump for $400\ \mathrm{ns}$, and when evaluating the final efficiency we pump for $10\ \mathrm{\mu s}$. 
%In the experiment, we use a cylindrical Pyrex cell wit dimensions 2.5cm x 7.5cm, filled with 5 Torr of $N_2$ buffer gas, kept at $60 \textdegree \rm{C}$, mounted inside a magnetic shielding. 
Differing to Ref. \cite{Esguerra.2023}, we operate the coherent signal pulses well above single photon level, at a continuous wave power of $126\ \mathrm{\mu W}$. Moreover, we detect the output of the memory directly on an amplified photodiode, that is, without further filtering. 
\\
A single memory experiment is illustrated in Fig.\ref{fig:1} c). The write pulse (purple) mediates the storage of the input signal pulse (blue) in a collective spin state of cesium atoms. After a time $\Delta t = 200 \rm{ns}$,  the read pulse (red) retrieves the stored optical field from the collective spin state; this output signal is referred to as the retrieved pulse. Any component of the optical field that is not stored in this process is referred to as the leakage. In this work we focus on learning the shape of the write pulse, the read pulse is fixed as a Gaussian with FWHM of $40\ \rm{ns}$. 
To quantify the performance of the memory, we define an internal memory efficiency  $\eta_{\mathrm{int}}$:

%we distinguish between the write pulse (purple), the shape of which we learn using the genetic algorithm, and the read pulse (red), a Gaussian pulse with  In the writing process, we send the signal field (blue) to be stored, and the write pulse into the atomic ensemble. we retrieve the optical field from the atomic coherence by sending in the control field.  The proportion of the input signal which we read out determines how well the memory performed.  We define an internal memory efficiency, $\eta_{int}$:

\begin{equation}\label{eq.1}
    \eta_{\rm{int}} = \frac{\int^{\Delta t+ \Delta t/2}_{\Delta t-\Delta t/2}|E_{\textrm{out}}|^2dt}{\int^{\Delta t/2}_{-\Delta t/2}|E_{\textrm{in}}|^2dt}.
\end{equation}
\\
%To measure $\eta_{\mathrm{int}}$ one must integrate over the retrieved optical pulse, and normalize against the pulse input into the system. 
$E_{\textrm{out}}$ is the retrieved signal pulse, $E_{\textrm{in}}$ is the initial signal and $\Delta t$ is the time between the initial signal and the retrieval control pulse. 
This input pulse, $|E_{\rm{in}}^{2}|$, is measured using the same detector, PD1 (Thorlabs APD0815), in the far-detuned regime, such that the optical losses in the system remain the same for both measurements, but the atoms do not absorb the signal. We red-detune the lasers by $\Delta = 2\ \mathrm{GHz}$ for this measurement. All the efficiencies of the final learned pulse are evaluated in this way.  The efficiencies in this work are reported for a storage time of $\Delta t = 200\ \rm{ns}$ and are not extrapolated back to $\Delta t = 0$ as would be standard in memory systems comparison. Consequently, we expect that we are under-representing the efficiencies by a factor of about 1.3.

To ensure consistency over multiple runs, before each evaluation of the internal efficiency, we turn off the heating of the atomic cell to further reduce stray magnetic fields and set the bias voltage of the EOM to its minimum.

%For other experimental details, we refer the reader to  \cite{Esguerra.2023}.
%Do I have to give details of the setup. 

\subsection{\label{sec:Genetic_alg}Genetic Algorithm}

The genetic algorithm \cite{Holland.1975} was developed in the 1960s and 1970s, inspired by considering how adaptation can be imported to computer systems, and thus to objective optimization \cite{Mitchell.1998}. The goal of a genetic algorithm is to iteratively find solutions which well fulfill the objective of a function.
 
\begin{figure}
    \centering
    \includegraphics[width = \linewidth]{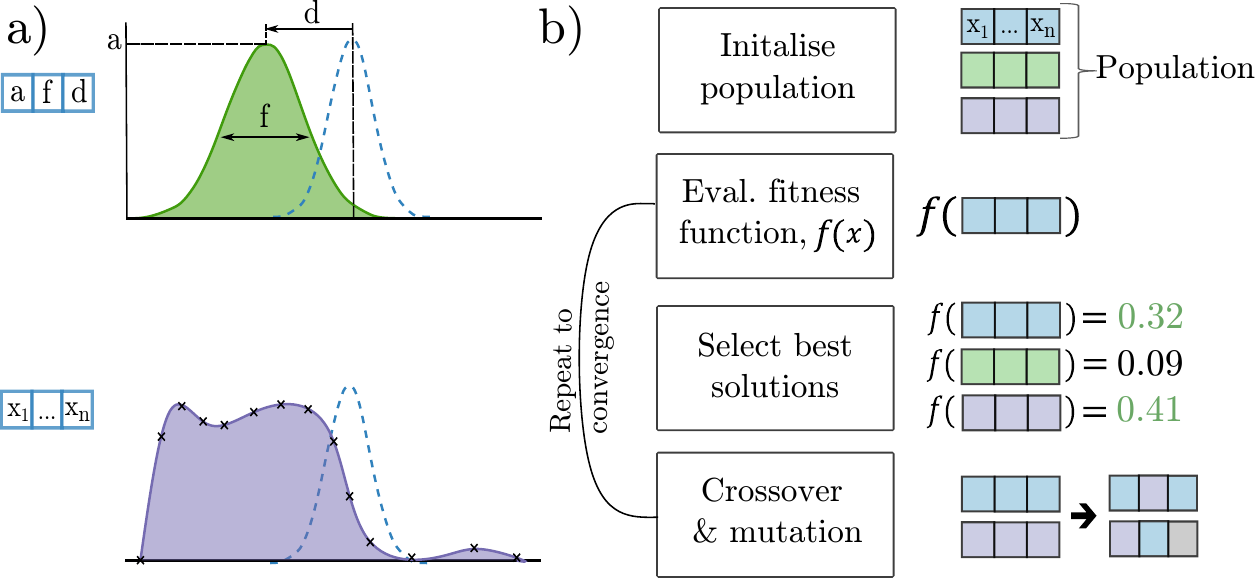}
    \caption{ a) Different gene representations for different solutions. Only 3 genes are used to encode the Gaussian; the amplitude, $a$, FWHM, $f$, and the delay with respect to the signal pulse, $d$. 16 evenly distributed points, $x_{\textrm{1}},...,x_{\textrm{16}}$ encode the free-form pulse. The signal pulse is shown in the dashed blue line. b) Overview of a genetic algorithm. }
    \label{fig:2}
\end{figure}

Figure \ref{fig:2} shows a flow chart of a genetic algorithm.  Briefly, solutions (individuals) are encoded using parameters (genes). Each solution is evaluated using an objective (fitness) function, which gives the optimization goal.  Solutions that have a high fitness, i.e. are appropriate solutions to the objective, are selected to form the next generation. The selected solution's genes are crossed over and mutated, generating the next generation's population. The genetic algorithm was implemented using the Python package PyGAD \cite{Gad.2021}.  

Here we wish to find the optimal control pulse shape, which maximizes the internal memory efficiency. Previously, Gaussians have been demonstrated as good  approximations of an ideal control pulse, and provide a good benchmark for performance analysis against a more general,  free-form pulse \cite{Esguerra.2023}\cite{Shinbrough.2023}. To be able to distinguish the improvement in efficiency due to the genetic algorithm from the improvement in efficiency due to the pulse encoding, we run the genetic algorithm for both Gaussian, and free-form solution representations, and compare the results. This corresponds to performing a constraint optimization like following:
\begin{equation}
    \hat{\theta} = \argmax_{\theta} \left( \eta_{\rm{int}}(\theta) \right) 
\end{equation}
where 
\begin{equation*}
    \theta =
    \begin{cases}
      [a,f,d] & \text{if encoding a Gaussian pulse}\\
      [x_1,...,x_{16} ] & \text{if encoding a free-form pulse}\\
    \end{cases}  
\end{equation*}
and  
 $\eta_{\rm{int}}$ is the efficiency of the memory experiment performed with a write pulse generated from $\theta$ (see Eq. \ref{eq.1}).
%Thus to fairly separate the improvement in effciency from the use of the genetic algorithm, and the inherent difference in improvement between guassian and free-form pulses, we apply a genetic algorithm to both solution representations and compare the results. 
%Two different representations of solutions were explored; solutions restricted to be exclusively gaussian in shape and solutions consisting of a 1D point cloud, interpolated to generate a continuous pulse form.

%For more details to the implementation of the genetic algorithm we refer the reader to appendix \ref{appendix:1}

In the Gaussian encoding, each solution is represented by three parameters: the amplitude of the Gaussian, $a$, the full-width-at-half-maximum of the Gaussian, $f$, and the delay of the pulse, $d$, with respect to the incoming signal pulse (see Fig. \ref{fig:2}a). The $a$ was allowed to take 50 discrete values normalized to between 0 and 1, to reduce the dimension of the feature space searched. Similarly, $d$ and $f$ were limited in their  possible range of values to between $-60$ and $0\ \rm{ns}$  and $(0$ and $80\ \rm{ns})$ respectively, both increasing $1\ \rm{ns}$ intervals. 

%Given the simplicity of the encoding, the system converged relatively quickly (See. Appendix 1. ) within an average time of 15 *FIND* epochs. As single epoch took approximately $\tilde 1.5 $ minutes, in the interest of reducing the time spent training our model, we limited the number of epochs to 25 before halting the algorithm.

To implement a free-form pulse optimization, we represent a solution by 16 evenly distributed points, which are interpolated using a smoothing cubic spline method, to form a continuous waveform. Each point is limited to a range of values between -0.2 and 1, and when interpolated, any negative values in the waveform are clipped to 0. This is to ensure that steep gradients can be learned, whilst not generating nonphysical negative pulse intensities. %Examples of genes (blue dashes) and the spline-generated pulse (blue continuous) can be seen in Figure \ref{fig:5}. 
\\
% Given the rise time of the AOM, the generated optical pulses vary slightly in shape, to the electrical waveforms.

For each of the two encodings, an initial population of 60 individuals is randomly generated. For each individual, we generate the electrical waveform encoded by the genes and modulate the waveform into the laser beam by altering the RF power driving the AOM. Then we perform a memory experiment, as detailed in Section \ref{sec:optMem}, using this waveform. The memory experiments for all individuals in a generation are recorded by an oscilloscope in a single shot and the trace is divided into individual memory experiments for evaluation. 

As mentioned in Section \ref{sec:optMem}, the internal efficiency is typically measured by detuning the laser, to ensure the signal pulse has the same losses as the retrieval pulse. However, to measure the signal far-detuned for each solution requires significant time overhead, which was deemed unfeasible for this experiment. Consequently, we evaluate the solutions on a fitness function which is faster to evaluate, and is linearly proportional to the internal efficiency. Specifically, to measure the fitness of each individual, we integrate the retrieved signal and normalize by a part of the input signal measured on a photodiode (PD2), before the cesium cell. The part of the input signal that is measured before the memory is integrated and is taken as a normalizing factor. 

\begin{figure}
    \centering
    \includegraphics[width = \linewidth]{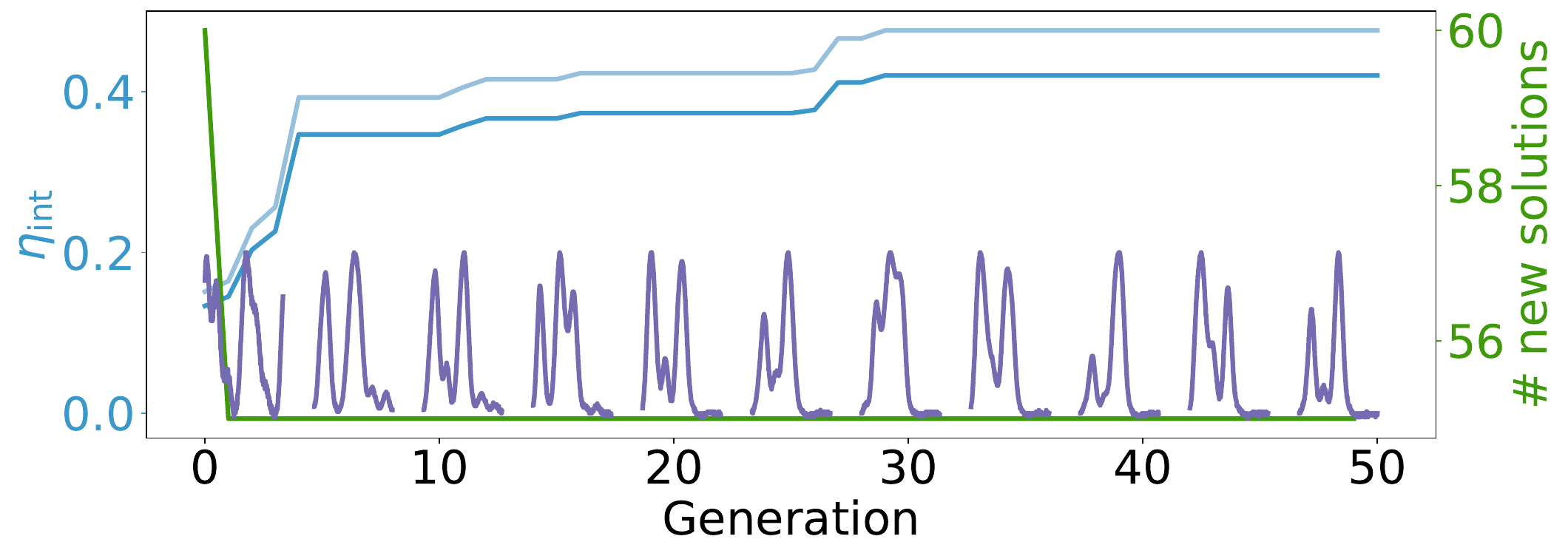}
    \caption{The convergence of the free-form pulse, learned for a 18 ns width signal. The fitness (light blue) and the related internal efficiency (dark blue) show the saturation of the learning process at generation 30. The pulse shapes for the best-performing pulse at each iteration are shown in purple, the final pulse is thus the end result. Green shows the number of new solutions tried per generation. }
    \label{fig:3}
\end{figure}

We select the solutions with the highest fitness to form the parents of next generation of solutions. This is done by tournament selection with a tournament size of ten, until ten parents are chosen. Once selected, the parents are crossed over uniformly, and each of the genes of the resulting children is randomly mutated with probability $p = 0.3$. 55 new children are generated in this manner and form the next generation, five of the best-performing solutions survive unchanged. This process is repeated for 50 (25) epochs for the free-form (Gaussian) pulses.  As a single evaluation of a generation takes $\sim 1.5$ minutes, the time taken to run the whole experiment is $1.5\ \rm{minutes} \times \rm{\#\ generations}$.  To ensure the timely termination of the algorithm, without cutting off the evolution too early, we fix the number of generations executed. Given the comparatively small feature space to be searched in the Gaussian case, we find the algorithm to converge before 25 generations.  Figure \ref{fig:3} shows a typical convergence plot for a single run of the genetic algorithm with free-form gene encoding, which illustrates the convergence of the algorithm well before 50 generations. Solutions that have been already evaluated are saved in a dictionary, and their value re-called if they are to be re-evaluated, resulting in strictly monotonous convergence.

%the retrieval normalised to part of the input signal, detected before the setup.
  %to generate a pseudo internal efficiency which we take as our fitness. We wish to maximise the internal efficiency, thus the higher the efficiency, the better performing the solution is. 
%specifically the AOM acts as a bandpass filter and high frequency components of the electrical pulse are filtered out. 

%Here we allow the system to evolve for 50 epochs, we find that the best solution is found on average at ... which the latest solution being found at.... 
%Do I need to give an example figure here? Should this go in the 
% I explictly do not explain these terms, because they are easily wikipediable. 

\section{ \label{sec: results} Results and discussion}
\label{sec:eff_opt}\subsection{Efficiency optimization}
\begin{figure}
    \centering
    \includegraphics[width = \linewidth]{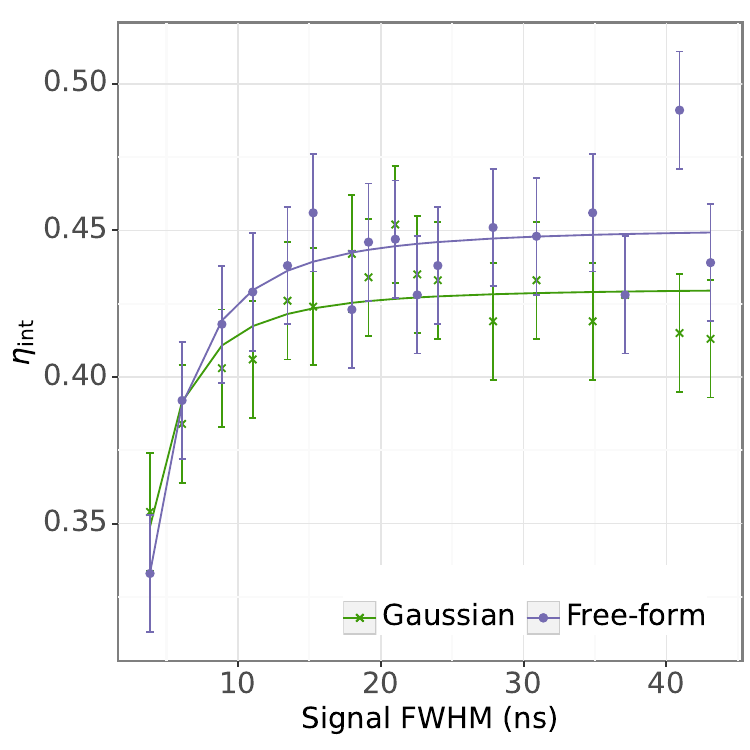}
    \caption{a) The efficiencies of the learned pulses, for varying signal FWHM. The free-from pulses (purple) look to perform on average, slightly better than the Gaussian pulse (green), but this improvement is not statistically significant. The fits are modelled as $\eta_{\rm{int}} = \eta_{0,\rm{t_S}}/ \sqrt{1 + (\frac{4 \ln(2)}{\Delta t_S \Delta \gamma })^2}$, where $\eta_{0,\rm{t_S}}$ is the maximal achievable efficiency and the bandwidth of the memory, $\gamma$. See Ref. \cite{Esguerra.2023} for the derivation.}
    \label{fig:4}
\end{figure}

We first run the genetic algorithm optimizing storage efficiency for different-width Gaussian input pulses. Here we scan the FWHM of the signal pulse from $3.8\ \rm{ns} -43\ \rm{ns} $ ($\tau_{\rm{FWHM}} \gamma  = \{1.9,21.5\}$, $\gamma \approx 500\ \mathrm{MHz} $), and for each signal, learn first the optimized Gaussian pulse, then the optimized free-form pulse. The results can be seen in Fig. \ref{fig:4}. On average, the free-from yields an improvement 3(7)\% over the learned Gaussian pulses. 
This extends the trend shown in Ref. \cite{Shinbrough.2021}, which shows that in a simulated three-level system, there is almost no difference in attainable efficiency between the free-form and Gaussian control pulses, for signal widths in the range  $\tau_{\rm{FWHM}} \gamma  = \{0.001,1.5\}$.  The agreement between simulation and experiment is particularly surprising, given the presence of known physical effects, which are not accounted for in three-level systems. Effects such as the inhomogeneous broadening of the excited state, differences in the Rabi frequency and the strength of coupling to different Zeeman levels are correlated to the pulse energy, and thus the temporal shape of the pulses used in the memory.  %One could expect that free-form pulses, by enabling a more varied control over the pulse energy, one could find an optimal timing for the peak strength rabi oscillation to be incident on the atoms, and thus increase the storage efficiency; this however does not seem to be the case 
%could find a pulse shape which is longer in duration, and thus with a higher total rabi frquency, typically coressponding to a higher storage efficiency; this however, does not seem to be the case.
\begin{figure*}[!ht]
    \centering
    \begin{subfigure}[b]{0.45\textwidth}
         \centering
         \includegraphics[width=\textwidth]{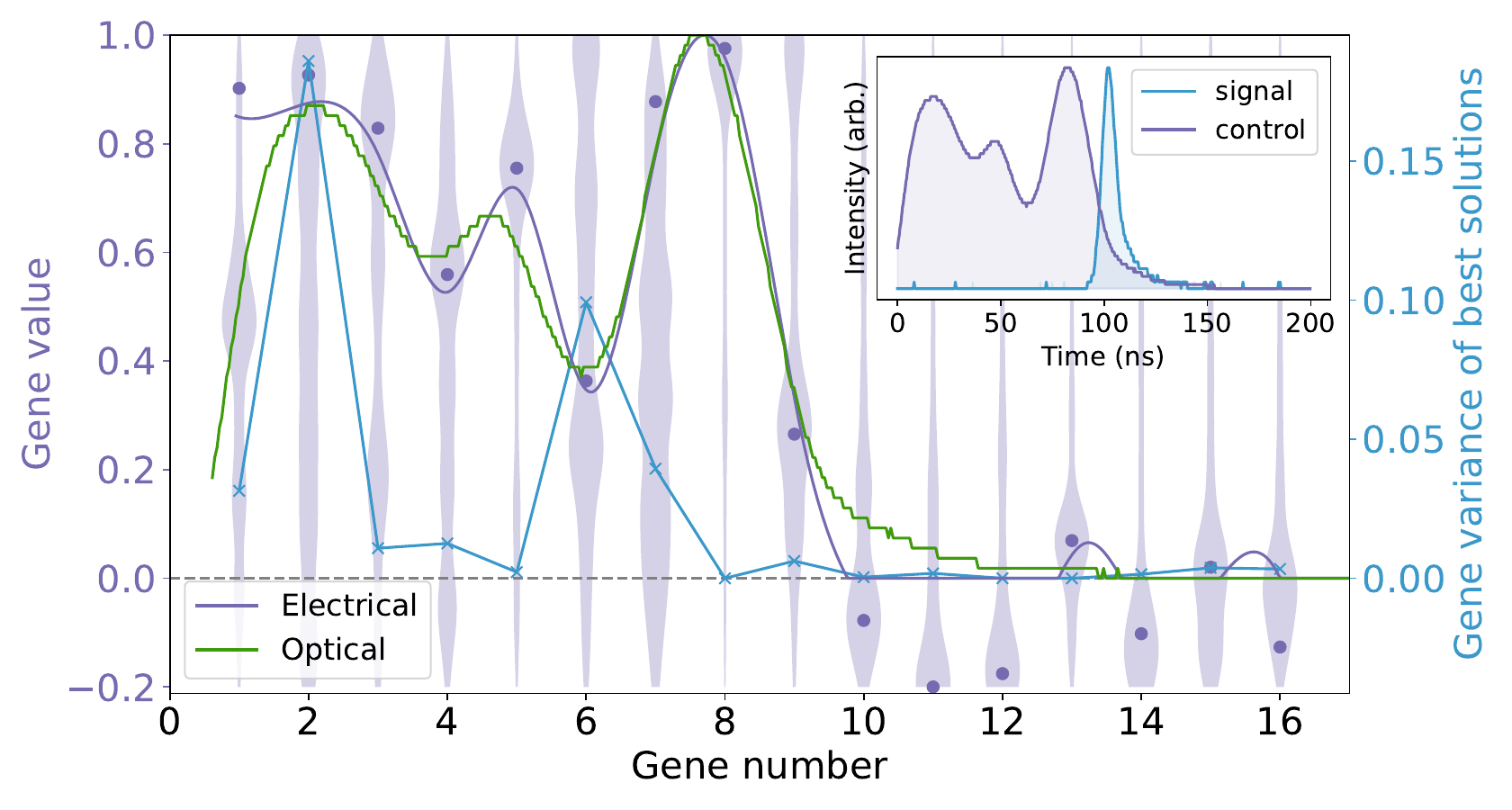}
         \caption{$8.9 \ \rm{ns}$ signal FWHM}
         \label{fig:10ns genes}
     \end{subfigure}
     \hfill
     \begin{subfigure}[b]{0.45\textwidth}
         \centering
         \includegraphics[width=\textwidth]{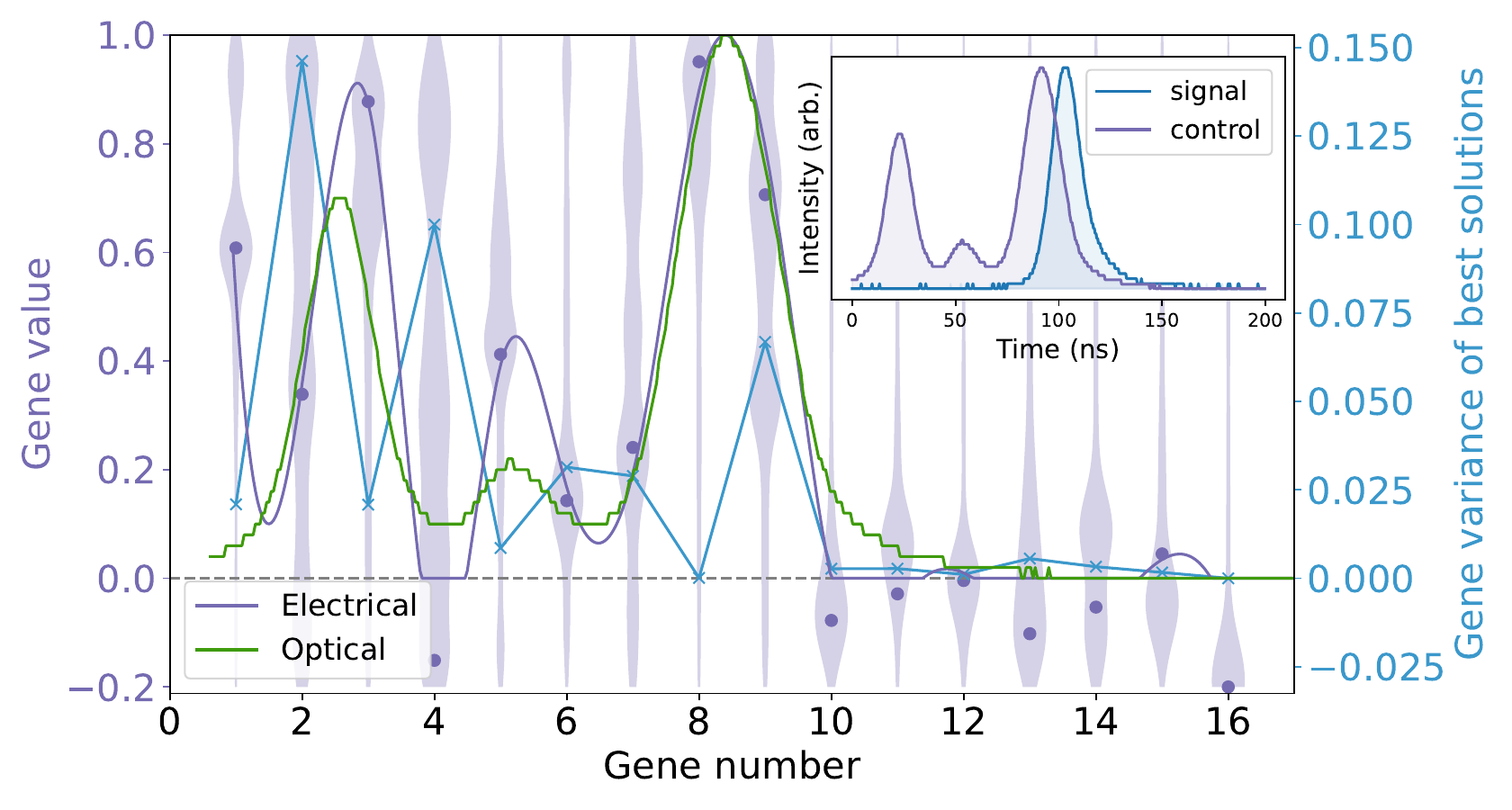}
         \caption{$18 \ \rm{ns}$ signal FWHM}
         \label{fig:20ns genes}
     \end{subfigure}
     \hfill
     \begin{subfigure}[b]{0.45\textwidth}
         \centering
         \includegraphics[width=\textwidth]{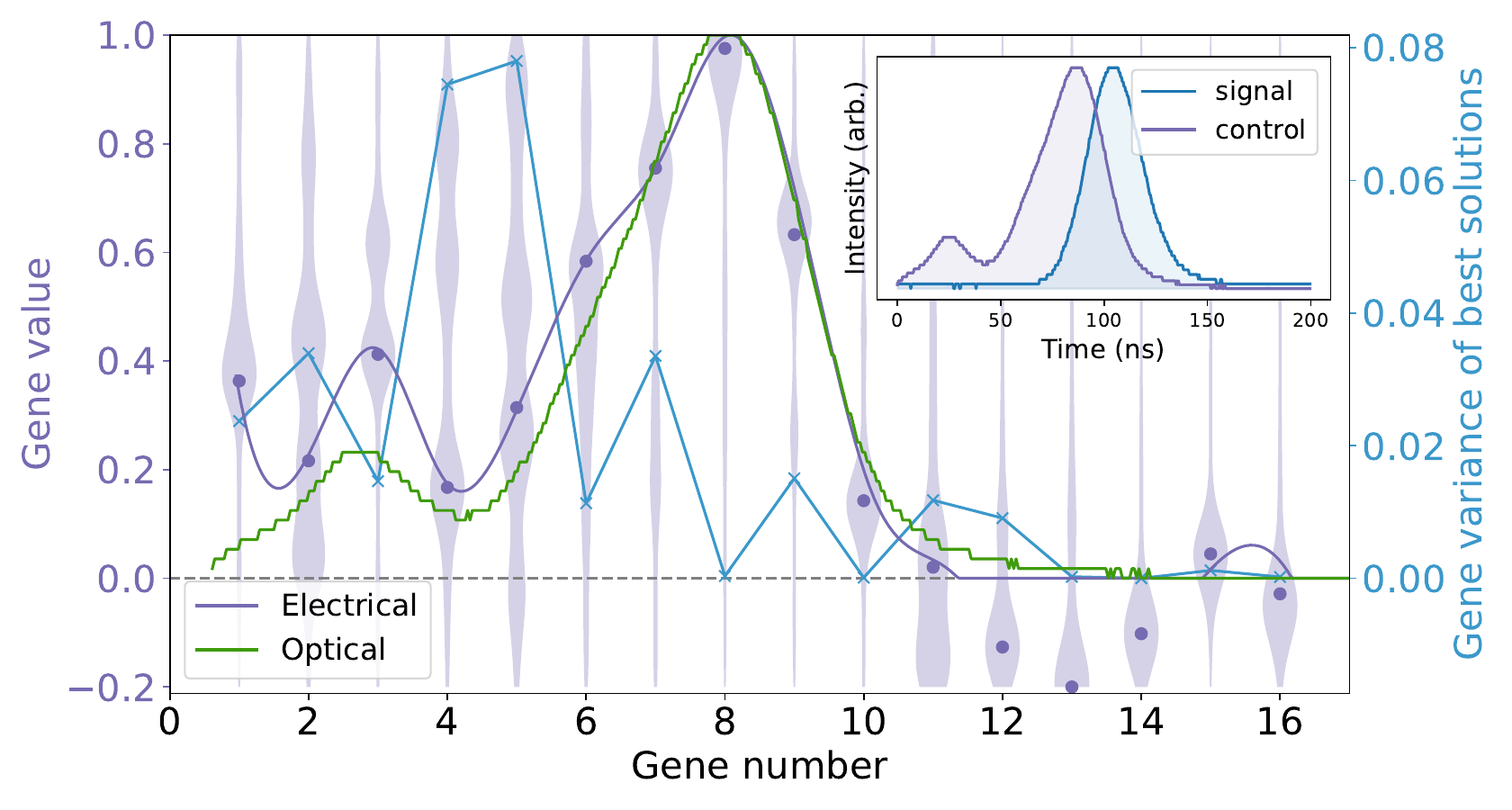}
         \caption{$31 \ \rm{ns}$ signal FWHM}
         \label{fig:40ns genes}
     \end{subfigure}
          \hfill
     \begin{subfigure}[b]{0.45\textwidth}
         \centering
         \includegraphics[width=\textwidth]{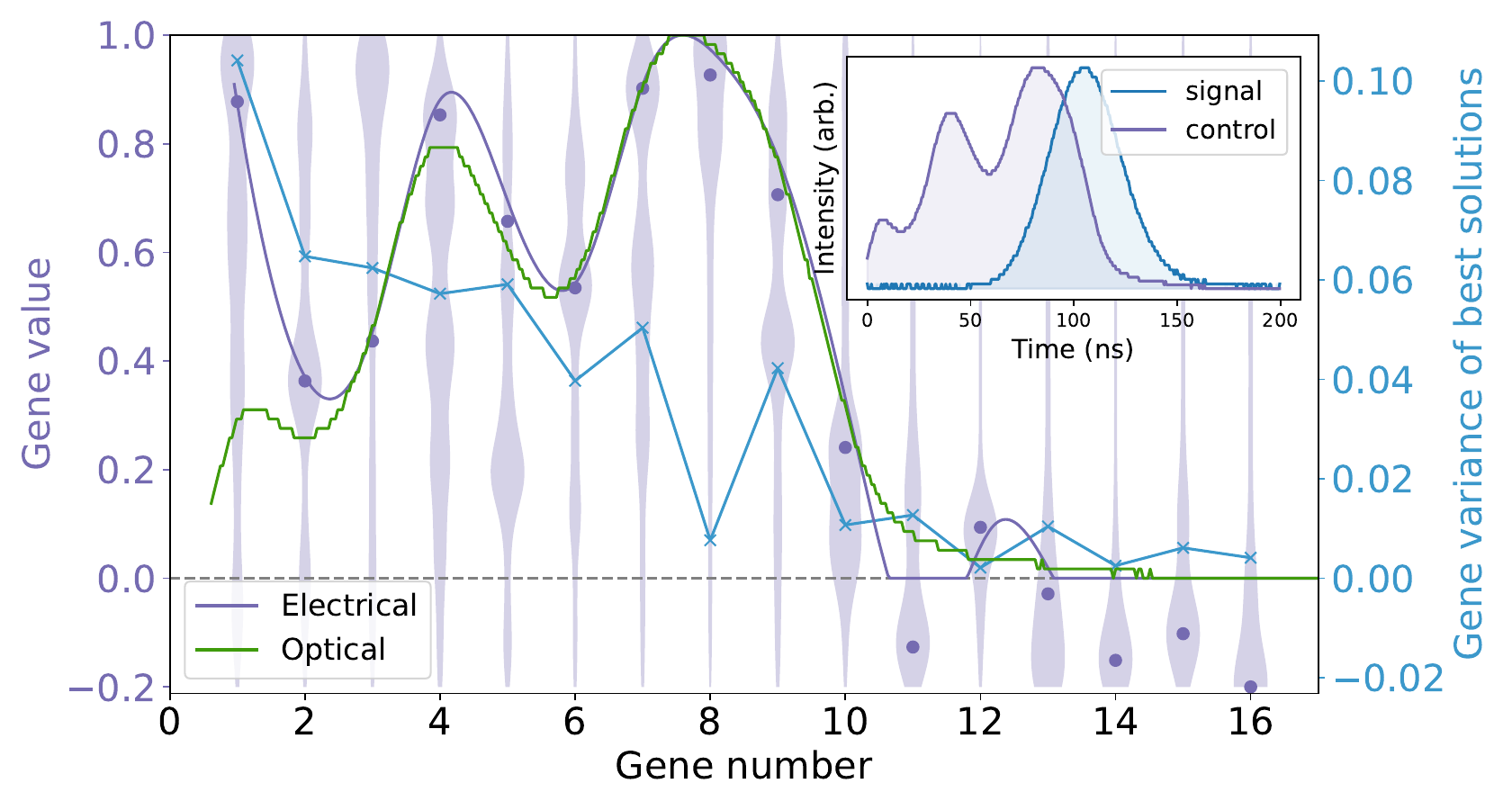}
         \caption{$41 \ \rm{ns}$ signal FWHM}
         \label{fig:55ns genes}
     \end{subfigure}
    \caption{Free-form pulse optimization. We show the best-learned pulses for a signal pulse with a) 8.9 ns b) 18 ns c) 31 ns and d) 41 ns full-width-at-half-maximum. The distribution of values explored for each gene throughout the whole optimization process is shown in a violin plot, with the final chosen gene values indicated as purple dots. The corresponding electrical waveform generated from the final gene values is shown in purple, and its corresponding optical signal in green. The variance of solutions that have a fitness within 10\% of the best-learned solution is plotted in blue. The lines joining the points have no physical significance and are intended as a guide to the eye. The temporal overlap of the signal and control pulses shows a consistent trend in the closing of the coherence window (inset). }
    \label{fig:5}
\end{figure*}

It seems as though the free-form pulses may yield only minimal improvements on the total storage efficiency, yet before accepting this conclusion one must also consider nuances of the pulse generation, particularly for smaller signal widths. Due to the $15\ \rm{ns}$ rise time of the AOM, for shorter signal pulses, it is not possible to modulate the optical field within the signal field for small signals. We are only able to modulate the field in the form of the falling flank of the AOM as it is switched off. Hence, it is not possible to definitively dismiss non-Gaussian pulses as non-optimal for signal pulses smaller than $\tau_{\rm{FWHM}} \gamma  = 1.9 $.
%This infers that one cannot alter the effects of FWM by cha
%Do we know that we have FWM in the system? yes - because luisa showed it in the noise.
%However, one must consider this conclusion with caution, given we anticipate that Gaussian signal pulses to perform well with Gaussian controls. Due to the long measurement times, an investigation of further pulse shapes was out of the scope of this work but is an area for further investigation.
%In the high-width pulse range, we know that Gaussians are a good approximation to optimal pulses. 
%This is also the range in which we have the bandwidth to modulate the pulse within the signal profile. For shorter pulses, the response of the AOM is too slow to modulate the control field within the signal field, so we cannot confirm that Gaussian pulses are optimal within this range. 
%Examples of the optimal free-form control pulses have been plotted in Figure \ref{fig:5}. Blue shows the electrical pulse, and green is the optical response of the AOM. Particularly noticeable in Figure \ref{fig:10ns genes}) is the $15\ \rm{ns}$  rise time of the AOM.

Figure \ref{fig:5} shows the final learned pulses for a subset of the signal widths. A general feature of all learned free-form pulses is a Gaussian-like falling edge temporal overlap with the signal pulse. For signal widths above 18ns, we observe a positively increasing trend between the distance of the peaks of the signal and control fields (see Fig. \ref{fig:5} insets).  Qualitatively, we see the downward slope of the control pulse crosses first with the rising edge of the signal pulse, i.e., the optimal control pulse arrives before the signal. This behavior is also consistent for the Gaussian pulses and is consistent with the optimal pulse learned in Ref. \cite{Shinbrough.2023}.

\begin{figure*}[!ht]
    \centering
    \includegraphics[width = 0.70\linewidth]{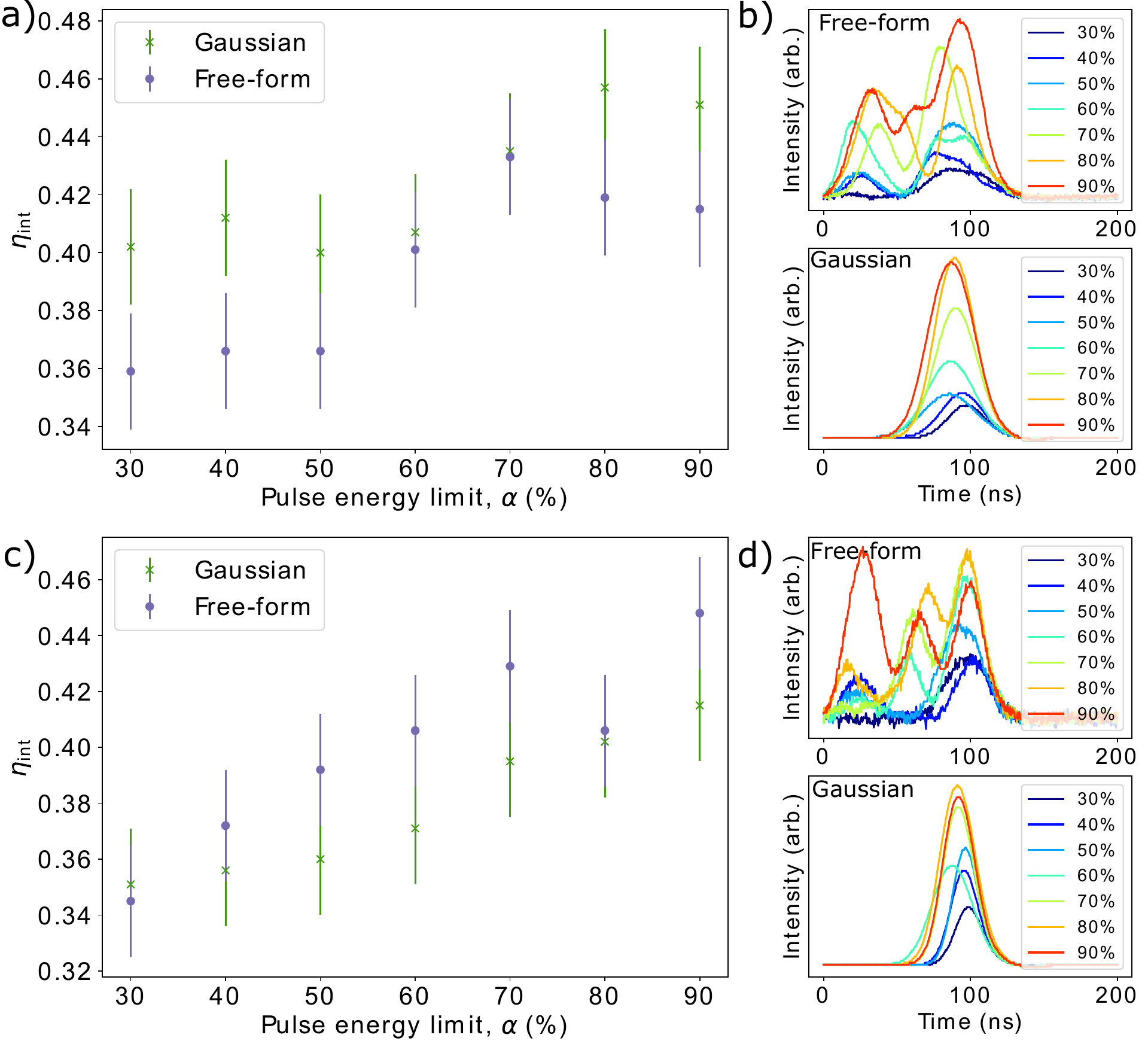}
    \caption{The efficiencies of energy restricted pulses, for a) 31 ns and c) 18 ns FWHM signal pulses. The corresponding learned pulses are plotted for the free-form pulses (plotted in the upper panel of b) and d)) and Gaussian pulses (plotted in the lower panel of b) and d)), with a color gradient from blue to red (energy: low - high).}
    \label{fig:6}
\end{figure*}

To determine how important specific genes are to the efficiency we use two indicators, the distribution of the state space explored during the learning process, and how much the genes of well-performing solutions vary.
First, we consider the state-space search by plotting the value distribution of all the solutions (purple violin plot in Fig.\ref{fig:5}). Genes where the importance of the value has little effect on the overall internal efficiency will have a wide value distribution, as seen in genes 2 and 7 of Figure \ref{fig:10ns genes}) and genes 2 and 4 of Figure \ref{fig:20ns genes}).  Similarly, genes whose value is important to the efficiency of the experiment will have a compact value distribution, such as most of the genes in the $11-16$ range of genes. This is not a surprise, as high gene values would trigger early read-out of the pulse, reducing the residual signal stored in the atoms, and thus reducing the retrieval.
Secondly, we calculate the variance for solutions that are at least 90\% of the maximum fitness. These can be seen in blue in Figure \ref{fig:5}.  We see a consistent trend across all learned pulses that the variance fluctuates in genes 1-9 and beyond genes 10 the fluctuations reduce, or are not present. This is an effect of preventing the aforementioned read-out of the pulse.  We see the variance mirrors the distributions over the whole searched space, that is, genes that vary over the whole state space, also have a high variance in the top 10\% of solutions. Thus, one can conclude that solutions that have a low variance are important to the efficiency, and thus should be the focus of further optimization efforts.  
%We see that several of the genes with a compact state space distribution (genes 13 and 15 in \ref{fig:10ns genes})) have a high variance. This does not necessarily show the lack of importance of the gene, rather is an artifact of the driving response of the AOM. In Fig. \ref{fig:55ns genes}) gene 13 has been set to a value above zero, but this is not present in the optical signal, as the change in RF amplitude was too small and too short to drive a response from the AOM. 
%Moreover, we can conclude that genes with a high distribution and a high variance, are less important to the efficiency. 
We find the concentration of high varying genes before the arrival of the signal pulse unsurprising, as these genes encode for time before the arrival of the signal pulse.

\subsection{ \label{sec:power_opt} Energy optimization}
One application we demonstrate in this report is to learn solutions limited in the total pulse energy. 

Ref. \cite{Esguerra.2023} shows the effect of pulse energy on the signal-to-noise ratio of the retrieved pulses from the optical memory. By setting genes of the pulse that have no effect on the efficiency to high values, one increases the total pulse energy. This, in turn,  inadvertently decreases the signal-to-noise ratio of the memory, with no efficiency payoff.  To mitigate this effect in solutions learned by the genetic algorithm, one can reduce the allowed energy through an energy optimization. Indeed, a benefit to using a genetic algorithm is that we can set secondary optimization objectives, to learn optimal pulses given particularly restrained conditions.

To optimize the energy, we set the maximally learned pulse energy to be the area under the curve of the electrical pulses learned in the experiments described in Sec.\ref{sec:eff_opt}. Then, we generate new solutions as described in Sec.\ref{sec:Genetic_alg}, but re-normalize any pulses whose electrical pulse area is larger than a specified limit, hereby setting a hard constraint on the allowed pulse energy of all the generated solutions. This leads to a modified constraint optimization function:

\begin{equation}
    \tilde{\theta} = \argmax_{\theta} \bigl( \eta_{\rm{int}}^{\prime}(\theta)  \bigr),
\end{equation}
where 
\begin{equation*}
\eta_{\rm{int}}^{\prime}  =
    \begin{cases}
      \eta_{\rm{int}}(\theta) & \text{if}\ I(\theta)\leq I(\hat{\theta})\alpha \\
      \eta_{\rm{int}}(\theta / \beta) & \text{otherwise.}\\
    \end{cases}
\end{equation*}
Here $I(\theta)$ is the integrated area under the curve of the pulse encoded by $\theta$, $\hat{\theta}$ are the learned parameters from the efficiency optimization, $\alpha$ is a percentage factor of the max energy, $I(\hat{\theta})$ and the normalizing factor, $\beta$ is given by $\beta = I(\theta) / I(\hat{\theta})\alpha $.
We choose the energy limit to be a progressively decreasing percentage of the learned pulse ($\alpha = \{0.3,0.9\}$), as shown in Figures \ref{fig:20ns genes}) and \ref{fig:40ns genes}). We carry out the energy optimization in the medium ($18\ \rm{ns}$) and long ($31\ \rm{ns}$) regimes and the results can be seen in Figure \ref{fig:6}.  The upper pulse energy limit corresponds to an optical pulse energy of  $E_{\rm{G},18  \rm{ns}} = 30.9\ \rm{nJ}$ and  $E_{\rm{G},31  \rm{ns}} = 33.0\ \rm{nJ}$  for the Gaussian pulses, and $E_{\rm{F},18  \rm{ns}} = 27.0\ \rm{nJ}$, $E_{\rm{F},31  \rm{ns}} = 52.2\ \rm{nJ}$ and for the free-form pulses. 
%$P_{\rm{F},8.9  \rm{ns}} = 66.0\ \rm{nJ}$,$P_{\rm{G},8.9  \rm{ns}} = 10.5\ \rm{nJ}$

We show that for a 31 ns signal pulse, we see a consistent trade-off in allowed pulse energy and attainable internal efficiency. This suggests that without the need for an extra optimization objective, the algorithm already goes part of the way to learning the most efficient pulse. Moreover, when we consider the absolute energy of the learned pulses, we note that the maximum allowed energy for the free-form pulses is about 1.6 times larger than the corresponding maximum Gaussian energy. This means, that in absolute terms, the 60\% energy free-form pulse, and 90\% Gaussian pulse have been limited to about the same energy, and yield similar efficiencies, correspondingly. This supports the conclusion that for large signal sizes, Gaussians are good approximations of the most energetically efficient pulse. Furthermore, we see in Fig. \ref{fig:6}b) that the falling flank of both the learned free-from and Gaussian pulses remains consistent as the energy is reduced, supporting the conclusions drawn from the previous section.
%learning were are able to consistently learn a free-from pulse with higher internal efficiency than their restricted Gaussian counterparts. Indeed we are able to get a 40\% reduction in the pulse power, and still achieve the same efficiency at the 

However,  for a $18\ \rm{ns}$ signal pulse (see Fig. \ref{fig:6}c), we are able to reduce the pulse energy of both the free-form and Gaussian pulses by 30\% with a reduction of only 4(6) \% from the efficiency at 100\% pulse energy.  For the free-from pulses, this reduction in energy corresponds to the reduction in the first part of the pulse (from 0- 75 ns), whilst the falling flank remains largely constant. This supports our conclusions from the gene analysis, that the first genes are not important factors in the efficiency. Indeed, we see that in the very low energy regime, we tend to a pulse shape that looks similar to the Gaussian. It is important to note, that only Figure \ref{fig:6}a) shows a signal pulse wide enough, such that one would be able to modulate the control pulse within the signal pulse, due to limitations in the AOM rise time.

%Figure \ref{fig:6} e) shows the efficiency in the short regime, at 8.9 ns signal FWHM. Here we see that the free-form pulses result in similar efficiencies across all the power levels, the Gaussian pulses decrease with the allowed power.  However, it is yet again important to note the discrepancy between the difference in absolute power, the Gaussian pulse limit is a sixth of the free-form limit, thus the Gaussian have a much lower energy limit and comparatively high performance. This discrepancy arises from differences in the best-learned pulses from the \textcolor{red}{...} optimization 

%Here we see that as we decrease the allowed power, we see the front flanks of the pulse reduce. 

%- The gradient of the downward slope stays constant for nearly all power optimizations.
While this method is not able to reduce the pulse energy whilst maintaining efficiency for all signal widths, it does give us a methodology that allows us to specify a pulse energy, decided by the desired signal-to-noise ratio, and learn optimal pulses for that energy level. This can be key in further pushing the efficiency of warm quantum memories.

\section{Conclusion}
In summary, we have shown that a genetic algorithm can be used to optimize the write pulses of an optical quantum memory. The choice of a free-form pulse encoding gives an overall average improvement factor of 3(7)\%, suggesting that Gaussian pulses are an acceptable approximation to optimal pulses. This agrees with the findings of Ref. \cite{Shinbrough.2023}. Nonetheless, we have demonstrated the merit of genetic algorithms in optimization with additional constraints, such as a total pulse energy limit. Here we demonstrated that it is possible to reduce the pulse energy by 30\% in some cases, without a large compromise in efficiency. This work focused on the measurement of the write control pulse optimization for Gaussian-shaped signal pulse, however, other works investigating signal shapes closer to the emission of single photon sources ( See Ref. \cite{Rakher.2013}), indicate the value in extending this approach to different signal shapes. Moreover, future work considering the learning and analysis of an optimal read pulse may be interesting to determine the validity of the time reversal assumption, often used in theoretical modeling \cite{Gorshkov.2007b}. Finally, this is a platform-agnostic approach that can be applied to a wide range of atomic and molecular physics experiments, supporting the further development of a range of high-efficiency optical quantum memories.
\section{Acknowledgments}
This work was funded by the German Ministry of Education and Research (BMBF) project Q-ToRX and Deutsche Forschungsgemeinschaft 448532670.
E.R. acknowledges funding through the Helmholtz Einstein International Berlin Research School in Data Science (HEIBRiDS).
\section{Appendix 1 - Genetic Algorithm details}
\label{appendix:1}

\begin{table}[h]
    \centering
    \begin{tabular}{l|l}
    Hyper-parameter &  Free-form (Gauss) \\
    \hline
       \# Genes  & 16 (3)  \\
       \# Generations & 50 (25) \\
       \# Solutions per population & 60 \\
       \# Parents mating & 10\\
       Selection type & Tournament \\
       Tournament size & 10 \\
       Elitism size  &  5 \\
       Crossover type & Uniform \\
       Mutation type & Random \\
       Mutation probability & 0.3 \\
    \end{tabular}
    \caption{The hyper-parameters used to run the genetic algorithm \cite{Gad.2021}. Values quoted in parenthesis are the parameters used for the Gaussian optimization.}
    \label{tab:hyper-param}
\end{table}

In this appendix we elaborate on the processes used to select the parents and generate the children of the next generation.  Once the fitness's of each solution have been evaluated, one must consider a method of choosing which solutions will be selected as parents. There are many different selection methods available; in this work, we chose tournament selection as it is easy to conceptualize and select the selection pressure (the likelihood that sub-optimal solutions are selected). In tournament selection, a subsection of the solutions are drawn at random. The highest-performing solution from the subset is selected as a parent for the next generation. This process is repeated until one has generated a number of parents, specified by the \# parents mating hyper-parameter. The list of parents is traversed sequentially and each pair of parents is used to generate a solution of the next generation by crossover, i.e. first parents 1 and 2 are selected, then parents 2, 3 etc. Once two parents have been selected and their genes are crossed over, there are several methods for performing crossover, which can be selected depending on the physical meaning of the genes. We select a uniform crossover, such that for each gene, one of the genes of the two parents is chosen with equal probability, and that gene is copied across to the other. Once crossed over, each gene of the child is mutated with a probability $p$, and the resulting solution is taken as one element of the population in the next generation.
Table \ref{tab:hyper-param} lists the hyper-parameters that were used in the experiment. 
\bibliography{lib}

\end{document}